\documentclass[aps,preprint,amsmath,amssymb,11pt]{revtex4}
\usepackage{graphicx}
\usepackage{epstopdf}
\pdfoutput=1 
\usepackage[T1]{fontenc}
\newcommand\sss{\scriptscriptstyle}
\newcommand{\mW}{m_{\sss W}}
\newcommand{\sW}{s_{\sss W}}
\newcommand{\cW}{c_{\sss W}}

\begin{document}

\title{Sensitivity reach on anomalous Higgs couplings via triphoton production for the post-LHC circular high-energy hadron colliders}
\author{H. Denizli}
\email[]{denizli_h@ibu.edu.tr}
\author{A. Senol}
\email[]{senol_a@ibu.edu.tr} 
\affiliation{Department of Physics, Bolu Abant Izzet Baysal University, 14280, Bolu, Turkey}
\begin{abstract}
The potential of triphoton production to obtain limits on anomalous Higgs boson couplings at $H\gamma\gamma$ and $HZ\gamma$ vertices is studied in Standard Model Effective Field Theory (EFT) framework for the post-LHC circular high-energy hadron colliders: High Luminosity-LHC (HL-LHC), High Energy LHC (HE-LHC) and Low Energy FCC (LE-FCC) which are designed with standard configurations of 14 TeV/3 ab$^{-1}$, 27 TeV/15 ab$^{-1}$ and 37.5 TeV/15 ab$^{-1}$. Madgraph in which the effective Lagrangian of the SM EFT is implemented using FeynRules and UFO framework is used to generate both background and signal events. These events are then passed through Pythia 8 for parton showering and Delphes to include realistic detector effects. After optimizing cuts on kinematics of three photons as well as the reconstructed invariant mass of the two leading photons, invariant mass of three leading photon is used to obtain constraints on the Wilson coefficients of dimension-six operators. We report the result of two dimensional scan of $\bar{c}_{\gamma}$ and $\tilde{c}_{\gamma}$ couplings at 95\% confidence level and compare with LHC results. Our obtained limits without systematic error on $\bar{c}_{\gamma}$ ( $\tilde{c}_{\gamma}$) are  $[-3.15;1.41]\times10^{-2}$ ($[-2.12;2.12]\times10^{-2}$), $[-1.21;0.78]\times10^{-2}$ ($[-0.98;0.98]\times10^{-2}$) and $[-0.89;0.66]\times10^{-2}$ ($[-0.77;0.77]\times10^{-2}$) for HL-LHC, HE-LHC and LE-FCC, respectively.

\end{abstract}

\maketitle

\section{Introduction}

Being the largest scientific instrument ever built, the Large Hadron Collider (LHC) with its discovery potential gave an opportunity to work in the physics of hadronic matter at extreme temperature and density for the large particle physics community. With the discovery of Higgs boson \cite{Aad:2012tfa, Chatrchyan:2012xdj}, Standard Model (SM) is completed and particle physics reach to the important moment of its history. Some of the important questions which are still not answered are; the nature of dark matter, the origin of the matter-antimatter asymmetry in the Universe, and the existence and hierarchy of neutrino masses. To address these questions and the properties of the newly discovered Higgs boson, Particle Physics community decided to upgrade LHC by increasing its luminosity (rate of collisions) by a factor of five beyond the original design value and the integrated luminosity (total collisions created) by a factor ten to sustain and extend its discovery potential. Possibility of building even higher energy frontier colliders is also under consideration. One can expect that these post-LHC circular high-energy hadron colliders will deepen our understanding of the origin of the electroweak symmetry breaking, Higgs couplings and new physics.

The HL-LHC program aims to decrease the statistical error in the measurements by half while it will continue to examine the properties of Higgs boson and look for clues to explain the physics beyond SM \cite{ApollinariG.:2017ojx}. The HL-LHC project includes a range of beam parameters and hardware configurations that will reach an integrated luminosity of approximately 250 fb$^{-1}$ per year after upgrading, achieving a target of 3000 fb$^{-1}$ at 7.0 TeV nominal beam energy reached by the LHC in a total of 12 years. One of the circular high energy hadron colliders under consideration after HL-LHC is HE-LHC which will extend the current LHC center of mass energy by almost a factor 2 and deliver an integrated luminosity of at least a factor of 3 larger than the HL-LHC \cite{Abada:2019ono}. It will use the existing LHC tunnel infrastructure and FCC-hh magnet technology, that is 16 Tesla dipole magnet. The other one is the LE-FCC \cite{Mangano:2681366}, which is thought to minimize the cost of a future circular hadron collider housed in the FCC 100 km tunnel with 6 Tesla dipole magnet. This leads to a center of mass energy 37.5 TeV. It plans to deliver an integrated luminosity of at least 10 ab$^{-1}$ during 20-year operation. 

Measuring precisely the Higgs couplings has a great potential to give us detailed information on the new physics beyond the SM. The Effective Field Theory (EFT) approach is widely used in the search for possible deviations from the predictions of the Standard Model. In the EFT framework \cite{Buchmuller:1985jz}, new physics contributions beyond the SM described by higher dimensional operators in an expansion. These operators are invariant under the SM symmetries and suppressed by the new physics scale $\Lambda$. In this article, we investigate potential for limitation of the EFT approach related to unitarity to describe possible contributions of the operators between Higgs and SM gauge boson at High-Luminosity LHC (HL-LHC) as well as other post-LHC hadron-hadron colliders under consideration. These operators are extensively studied via different production mechanism for hadron colliders \cite{Hagiwara:1993qt,Corbett:2012ja,Ellis:2014jta,Ellis:2014dva,Falkowski:2015fla,Corbett:2015ksa,Ferreira:2016jea,Aad:2015tna,Englert:2015hrx,Buckley:2015lku,Englert:2016hvy,Degrande:2016dqg,Bishara:2016kjn,Liu-Sheng:2017pxk,Aaboud:2018xdt,Khachatryan:2014kca,Khachatryan:2016tnr,Khanpour:2017inb,vonBuddenbrock:2017gvy,Shi:2018lqf,Freitas:2019hbk,Aaboud:2019nan,Banerjee:2019jys}.

One of the approach would be to look for production which is rare in the SM. One production mechanism satisfying this in the hadron colliders is events with three photons in the final state \cite{deCampos:1998xx,GonzalezGarcia:1999fq,Aaboud:2017lxm}. This process also involves pure electroweak interactions at tree level. Therefore, the production mechanism stands out as an ideal platform to search for deviations from SM. In the literature, the three photon final state as well as other mechanism \cite{Achard:2004kn,Abreu:1999vt,Heister:2002ub,Denizli:2019ijf} are used in the search of anomalous Higgs couplings via EFT formalism. Either direct production or fragmentation process result in three photon final state in the hadron colliders. Since photons produced via direct production are typically isolated, requiring isolated photons will reduce the background contributions from the decays  of unstable particles such as $\pi^0\to \gamma\gamma$ and suppress the signal process with one or more fragmentation photons.

The organization of the paper is as follows. We highlight some details of the model that
are relevant to our study and calculate the cross-sections as a function of couplings under consideration for each post-LHC hadron colliders in Section II. Discussion of kinematic cuts and the details of signal-and background analysis are given in Section III. In Section IV, we present obtained the sensitivity bounds on the $\bar{c}_{\gamma}$ and $\tilde{c}_{\gamma}$ couplings with two dimensional scan at 95\% confidence level and compare with LHC results. Finally, we conclude in Section V.
\section{Theoretical framework of the Effective Operators}
In this study, we are interested  the CP-conserving and CP-violating dimension-6 operators of the Higgs boson and electroweak gauge boson in the convention of the Strongly Interacting Light Higgs (SILH) basis in effective Lagrangian \cite{Alloul:2013naa}.
The CP-conserving dimension-6 operator between the Higgs boson and electroweak gauge bosons is defined in the general effective Lagrangian as follows
\begin{eqnarray}\label{CPC}
	\begin{split}
		\mathcal{L}_{\rm CPC} = & \
		 \frac{\bar c_{H}}{2 v^2} \partial^\mu\big[\Phi^\dag \Phi\big] \partial_\mu \big[ \Phi^\dagger \Phi \big]
		+ \frac{\bar c_{T}}{2 v^2} \big[ \Phi^\dag {\overleftrightarrow{D}}^\mu \Phi \big] \big[ \Phi^\dag {\overleftrightarrow{D}}_\mu \Phi \big] - \frac{\bar c_{6} \lambda}{v^2} \big[\Phi^\dag \Phi \big]^3
		\\
		& \
		  - \bigg[\frac{\bar c_{u}}{v^2} y_u \Phi^\dag \Phi\ \Phi^\dag\cdot{\bar Q}_L u_R
		  + \frac{\bar c_{d}}{v^2} y_d \Phi^\dag \Phi\ \Phi {\bar Q}_L d_R
		+\frac{\bar c_{l}}{v^2} y_l \Phi^\dag \Phi\ \Phi {\bar L}_L e_R
		 + {\rm h.c.} \bigg]
		\\
		&\
		 + \frac{i g\ \bar  c_{W}}{m_{W}^2} \big[ \Phi^\dag T_{2k} \overleftrightarrow{D}^\mu \Phi \big]  D^\nu  W_{\mu \nu}^k + \frac{i g'\ \bar c_{B}}{2 m_{W}^2} \big[\Phi^\dag \overleftrightarrow{D}^\mu \Phi \big] \partial^\nu  B_{\mu \nu} \\
		&\   
		+ \frac{2 i g\ \bar c_{HW}}{m_{W}^2} \big[D^\mu \Phi^\dag T_{2k} D^\nu \Phi\big] W_{\mu \nu}^k  
		+ \frac{i g'\ \bar c_{HB}}{m_{W}^2}  \big[D^\mu \Phi^\dag D^\nu \Phi\big] B_{\mu \nu}   \\
		&\
		 +\frac{g'^2\ \bar c_{\gamma}}{m_{W}^2} \Phi^\dag \Phi B_{\mu\nu} B^{\mu\nu}  
		+\frac{g_s^2\ \bar c_{g}}{m_{W}^2} \Phi^\dag \Phi G_{\mu\nu}^a G_a^{\mu\nu} 
	\end{split}
\end{eqnarray}
where $\lambda$ denotes the Higgs quartic coupling  and $\Phi$ represents the Higgs sector containing a single $SU(2)_L$ doublet of fields;  $g_s$, $g$  and $g'$ are coupling constant of  $SU(3)_C$, $SU(2)_L$  and $U(1)_Y$ gauge fields, respectively; the generators of $SU(2)_L$ in the fundamental representation are given by $T_{2k}=\sigma_k/2$ (here $\sigma_k$ are the Pauli matrices); $y_u$, $y_d$ and $y_l$ are the $3\times3$ Yukawa coupling matrices in flavor space; $\overleftrightarrow{D}_\mu$ correspond to the Hermitian derivative operators; $B^{\mu\nu}$, $W^{\mu \nu}$ and $G^{\mu\nu}$ are the electroweak and the strong field strength tensors, respectively.  

 The general effective Lagrangian can be extend with $CP$-violating operators in the SILH basis given below
 \begin{eqnarray}
\label{CPV}
  {\cal L}_{CPV} = &\
    \frac{i g\ \tilde c_{ HW}}{\mW^2}  D^\mu \Phi^\dag T_{2k} D^\nu \Phi {\widetilde W}_{\mu \nu}^k
  + \frac{i g'\ \tilde c_{ HB}}{\mW^2} D^\mu \Phi^\dag D^\nu \Phi {\widetilde B}_{\mu \nu}
  + \frac{g'^2\  \tilde c_{ \gamma}}{\mW^2} \Phi^\dag \Phi B_{\mu\nu} {\widetilde B}^{\mu\nu}\\
 &\
  +\!  \frac{g_s^2\ \tilde c_{ g}}{\mW^2}      \Phi^\dag \Phi G_{\mu\nu}^a {\widetilde G}^{\mu\nu}_a
  \!+\!  \frac{g^3\ \tilde c_{3W}}{\mW^2} \epsilon_{ijk} W_{\mu\nu}^i W^\nu{}^j_\rho {\widetilde W}^{\rho\mu k}
  \!+\!  \frac{g_s^3\ \tilde c_{ 3G}}{\mW^2} f_{abc} G_{\mu\nu}^a G^\nu{}^b_\rho {\widetilde G}^{\rho\mu c} \ \nonumber
\end{eqnarray}
where \begin{eqnarray}
  \widetilde B_{\mu\nu} = \frac12 \epsilon_{\mu\nu\rho\sigma} B^{\rho\sigma} \ , \quad
  \widetilde W_{\mu\nu}^k = \frac12 \epsilon_{\mu\nu\rho\sigma} W^{\rho\sigma k} \ , \quad
  \widetilde G_{\mu\nu}^a = \frac12 \epsilon_{\mu\nu\rho\sigma} G^{\rho\sigma a} \ \nonumber 
\end{eqnarray}
are the dual field strength tensors.

The dimension-6 CP-conserving (Eq.\ref{CPC}) and CP-violating (Eq.\ref{CPV}) operators in SILH bases can be defined in terms of the mass eigenstates after electroweak symmetry breaking. The relevant Higgs and neutral gauge boson couplings in the mass basis for triphoton production are given in following Lagrangian
\begin{eqnarray}\label{massb}
  {\cal L} &= &\ 
    - \frac{1}{4} g_{\sss h\gamma\gamma} F_{\mu\nu} F^{\mu\nu} h
    - \frac{1}{4} \tilde g_{\sss h\gamma\gamma} F_{\mu\nu} \tilde F^{\mu\nu} h
\nonumber\\
    &-& \frac{1}{4} g_{\sss hzz}^{(1)} Z_{\mu\nu} Z^{\mu\nu} h
    - g_{\sss hzz}^{(2)} Z_\nu \partial_\mu Z^{\mu\nu} h
    + \frac{1}{2} g_{\sss hzz}^{(3)} Z_\mu Z^\mu h
    - \frac{1}{4} \tilde g_{\sss hzz} Z_{\mu\nu} \tilde Z^{\mu\nu} h
\\ 
    &-& \frac{1}{2} g_{\sss haz}^{(1)} Z_{\mu\nu} F^{\mu\nu} h
    - \frac{1}{2} \tilde g_{\sss haz} Z_{\mu\nu} \tilde F^{\mu\nu} h
    - g_{\sss haz}^{(2)} Z_\nu \partial_\mu F^{\mu\nu} h 
    \nonumber
\end{eqnarray}
where the field strength tensors of $Z$-boson and photon are represented with $Z_{\mu\nu}$ and $F_{\mu\nu}$, respectively. The relationships between the effective couplings in the gauge basis and dimension-6 operators are given in Table~\ref{mtable}  in which $a_{H}$ coupling is the SM contribution to the $H\gamma\gamma$ vertex at loop level.

\begin{figure}
\includegraphics[scale=0.6]{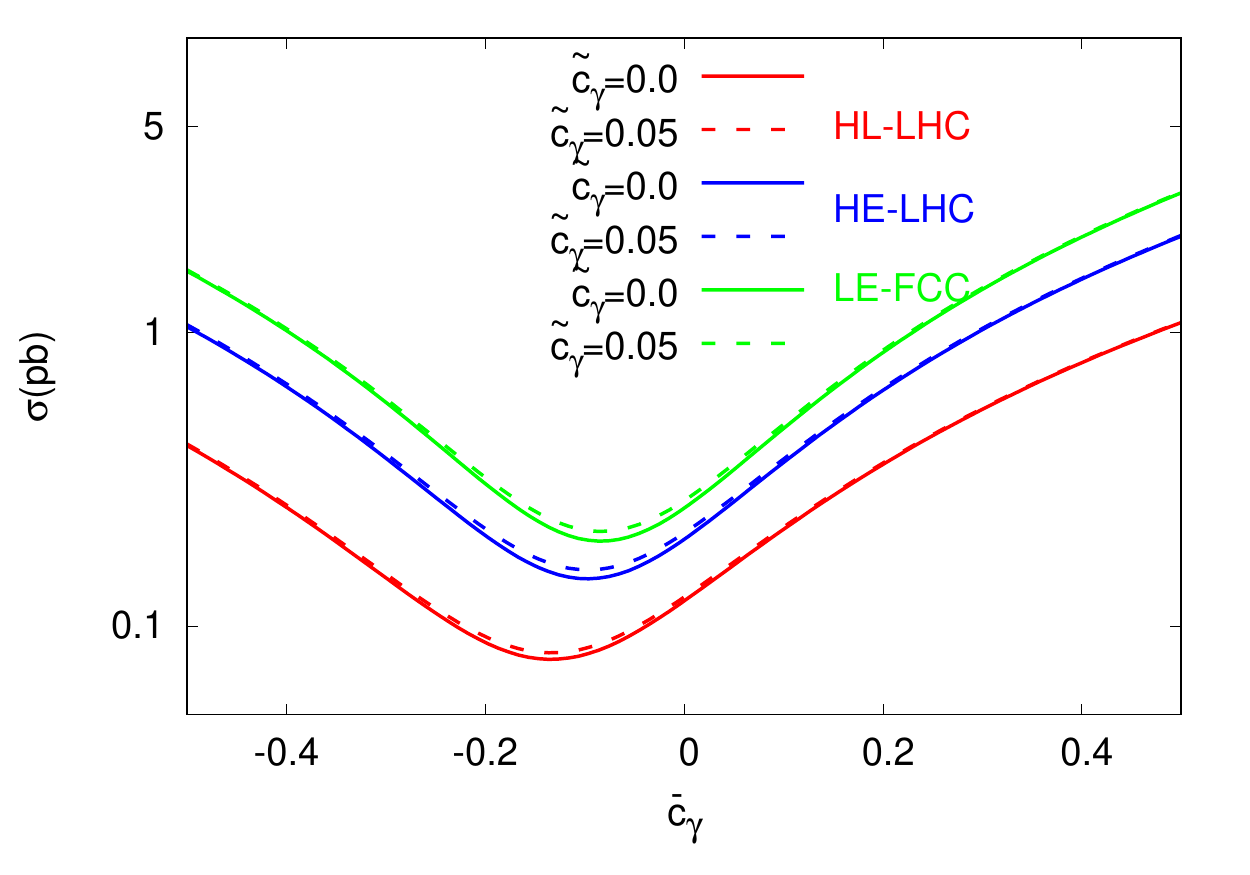}
\includegraphics[scale=0.6]{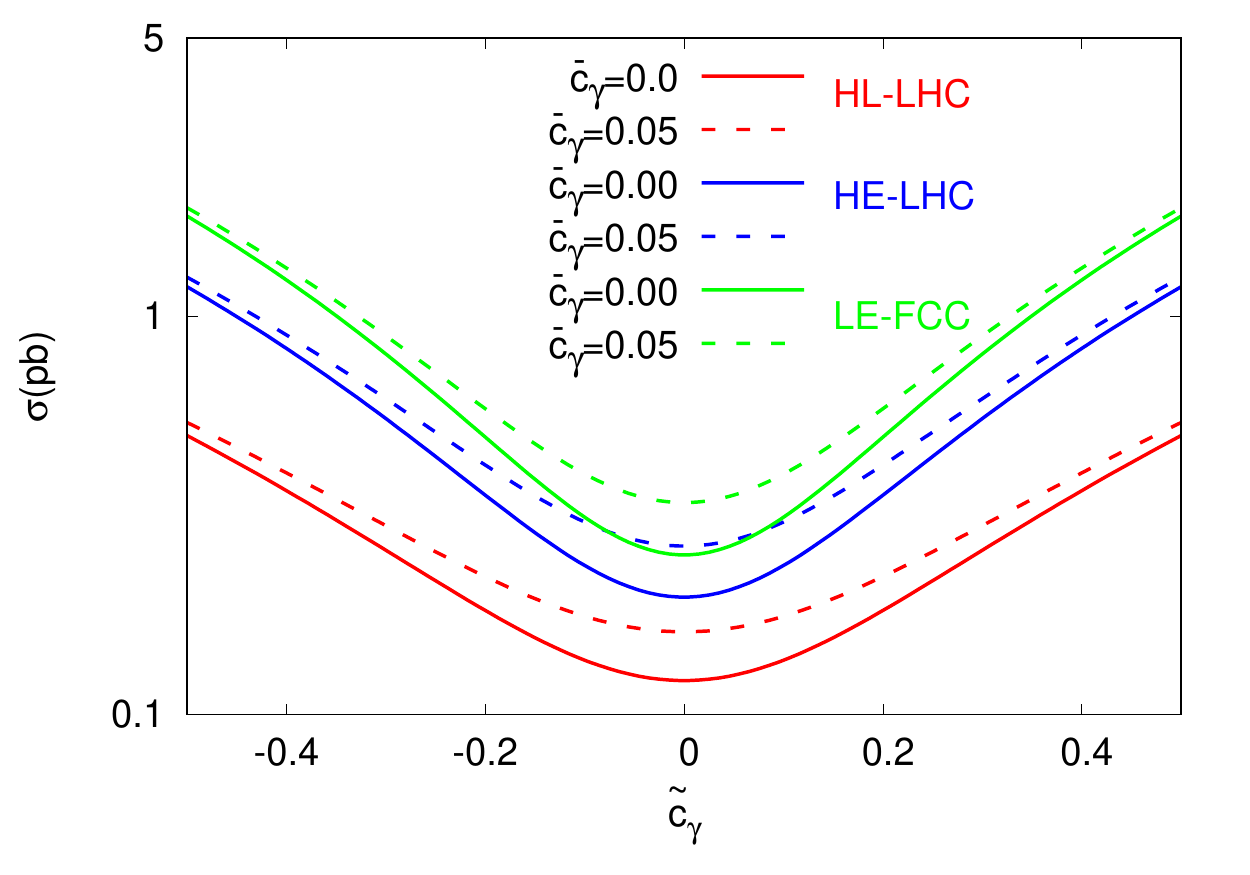}  
\caption{The total cross section as a function of the CP-conserving  $\bar{c}_{\gamma}$ couplings for $\tilde{c}_{\gamma}$=0 and 0.05 (left) and CP-violating $\tilde{c}_{\gamma}$ couplings (right) for $\bar{c}_{\gamma}$=0 and 0.05 $pp\to \gamma\gamma\gamma$ subprocess at the three post LHC circular colliders. \label{crosssection}} 
\end{figure}

\begin{table}[h]
\caption{The relations between Lagrangian parameters in the mass basis (Eq.\ref{massb}) and the Lagrangian in gauge  basis (Eqs. \ref{CPC} and \ref{CPV}). ($c_W\equiv\cos \theta_W$, $s_W\equiv\sin \theta_W$)}  
\begin{ruledtabular}\label{mtable}
\begin{tabular}{ll}
  $g_{h\gamma\gamma}$= $a_{ H} - \frac{8 g \bar c_{ \gamma} \sW^2}{\mW}$ & $\tilde g_{ h\gamma\gamma}$  
     $= -\frac{8 g \tilde c_{ \gamma} \sW^2}{\mW}$ \\
    $g^{(1)}_{ hzz}$= $\frac{2 g}{\cW^2 \mW} \Big[ \bar c_{HB} \sW^2 - 4 \bar c_{ \gamma} \sW^4 + \cW^2 \bar c_{ HW}\Big]$& $g^{(2)}_{ hzz}$= $\frac{g}{\cW^2 \mW} \Big[(\bar c_{ HW} +\bar c_{ W}) \cW^2  + (\bar c_{ B} + \bar c_{ HB}) \sW^2 \Big]$ \\
    $g^{(3)}_{hzz}$=  $\frac{g \mW}{\cW^2} \Big[ 1 -\frac12 \bar c_{H} - 2 \bar c_{T} +8 \bar c_{\gamma} \frac{\sW^4}{\cW^2}  \Big]$& $\tilde g_{ hzz}$ =$\frac{2 g}{\cW^2 \mW} \Big[ \tilde c_{ HB} \sW^2 - 4 \tilde c_{ \gamma} \sW^4 + \cW^2 \tilde c_{ HW}\Big]$ \\
     $g^{(1)}_{ h\gamma z}$= $\frac{g \sW}{\cW \mW} \Big[  \bar c_{ HW} - \bar c_{HB} + 8 \bar c_{ \gamma} \sW^2\Big]$& $\tilde g_{ h\gamma z}$ =$\frac{g \sW}{\cW \mW} \Big[  \tilde c_{HW} - \tilde c_{ HB} + 8 \tilde c_{ \gamma} \sW^2\Big]$ \\
    $g^{(2)}_{h\gamma z}$= $\frac{g \sW}{\cW \mW} \Big[  \bar c_{HW} - \bar c_{ HB} - \bar c_{ B} + \bar c_{ W}\Big]$&   \\
     \end{tabular}
\end{ruledtabular}
\end{table}
In order to simulate events involving the effect of the dimension-6 operators on the triphoton production mechanism in $pp$ collisions with leading order, the effective Lagrangian of the SM EFT in Eq.(\ref{massb}) is implemented into the \verb|MadGraph5_aMC@NLO v2.6.3.2| \cite{Alwall:2014hca} event generator using FeynRules \cite{Alloul:2013bka} and UFO \cite{Degrande:2011ua} framework. The triphoton production process is sensitive to the Higgs-electroweak gauge boson couplings ($g_{h\gamma\gamma}$ and $g_{hz\gamma}$) and the couplings of a quark pair to single Higgs field ( $\tilde y_u$ and $\tilde y_d$) in the the mass basis effective Lagrangian, the eight Wilson coefficients ( $\bar c_{ W}$,  $\bar c_{ B}$, $\bar c_{ HW}$, $\bar c_{ HB}$, $\bar c_{ \gamma}$, $\tilde{c}_{HW}$, $\tilde{c}_{HB}$ and $\tilde{c}_{\gamma}$) related to Higgs-gauge boson couplings and also effective fermionic couplings in the gauge basis effective Lagrangian.
Since Yukawa coupling of the first and second generation fermions is very small, the effective fermionic couplings are ignored. The coupling constants other than $\bar c_{ \gamma}$ and $\tilde{c}_{\gamma}$  couplings do not leads to considerable modifications in the cross section as seen in our previous work \cite{Denizli:2019ijf}. Thus we focus on the effect of $\bar c_{ \gamma}$ and $\tilde{c}_{\gamma}$  couplings on the triphoton production process in this study. We generate 64 samples to parametrize the cross section function by varying two Wilson coefficients simultaneously for HL-LHC as well as other post-LHC hadron-hadron colliders under consideration. For the studies presented in this manuscript, we assume $\sqrt s= 14$ TeV with $L_{int}$=3 ab$^{-1}$ for HL-LHC,  $\sqrt s= 27$ TeV with $L_{int}$=15 ab$^{-1}$ for HE-LHC and $\sqrt s= 37.5$ TeV with $L_{int}$=15 ab$^{-1}$ for LE-FCC as indicated in the Ref.\cite{Mangano:2681366}. We apply generator level cuts; $p_T^{\gamma_1,\gamma_2, \gamma_3} > 15$ GeV and $|\eta^{\gamma_{1,2,3}}| < 2.5$ in the calculations of the cross sections at the leading order. Then the method is validated by comparing the cross sections obtained with the parametrisation function to the cross section obtained from \verb|MadGraph5_aMC@NLO v2.6.3.2| with specific values of the couplings. Fig.\ref{crosssection} depicted the total cross section of $pp\to \gamma\gamma\gamma$ process as a function of the CP-conserving  $\bar{c}_{\gamma}$ couplings for $\tilde{c}_{\gamma}$=0 and 0.05 on the right panel and CP-violating $\tilde{c}_{\gamma}$ couplings  for $\bar{c}_{\gamma}$=0 and 0.05 on the left panel at the three post LHC circular colliders.
 In this figure, all effective couplings other than $\bar{c}_{\gamma}$ and $\tilde{c}_{\gamma}$  are set to zero. The effects of the $\tilde{c}_{\gamma}$ is smaller in the cross section as a function of $\bar{c}_{\gamma}$ in the right panel of Fig.\ref{crosssection} while $\bar{c}_{\gamma}$ contributions in cross section is significant in the right panel of Fig.\ref{crosssection} in range of the small coupling value of $\tilde{c}_{\gamma}$.

 \section{Signal and Background Analysis}
 
 In this section, we give details of the simulation and the cut-based analysis steps to explore the potential of triphoton production to obtain limits on anomalous Higgs boson $\bar{c}_{\gamma}$ and $\tilde{c}_{\gamma}$ couplings at $H\gamma\gamma$ and $HZ\gamma$ vertices in a model-independent Standard Model effective field theory framework for the post-LHC circular high-energy hadron colliders. This final state consists of one energetic photon together with two photons originating from the Higgs boson decay. Therefore, $pp \to\gamma\gamma\gamma$ process with non-zero $\bar{c}_{\gamma}$ and $\tilde{c}_{\gamma}$ effective couplings is considered as signal  including SM contribution as well as interference between effective couplings and SM contributions ($SB_{\gamma\gamma\gamma}$). The main sources of the SM background processes which are taken into account in this work are $pp \to\gamma\gamma\gamma$ ($B_{\gamma\gamma \gamma}$:  the same final state as the signal process) and $pp\to \gamma\gamma$+jet ($B_{\gamma\gamma j}$: in which jet may fake a photon). \verb|MadGraph5_aMC@NLO v2.6.3.2| event generator is used to generate 500 k events for the signal and background processes at leading order partonic level. The total of 64 samples for each  post-LHC hadron collider consideration is generated by varying two Wilson coefficients $\bar{c}_{\gamma}$ and $\tilde{c}_{\gamma}$  simultaneously. Consequently, these events are passed through the Pythia 8 \cite{Sjostrand:2014zea} including initial and final parton shower and the fragmentation of partons into hadron. The detector responses are taken into account with the card prepared for HL-LHC and HE-LHC studies
 released in \verb|Delphes 3.4.1| package \cite{deFavereau:2013fsa}. Jets in all generated events are clustered by using FastJet \cite{Cacciari:2011ma} with anti-$k_t$ algorithm where a cone radius is set as $\Delta R$ = 0.4 \cite{Cacciari:2008gp}. All events are analysed by using the ExRootAnalysis utility \cite{exroot} with ROOT \cite{Brun:1997pa}.  
 \begin{figure}[htb!]
\includegraphics[scale=0.8]{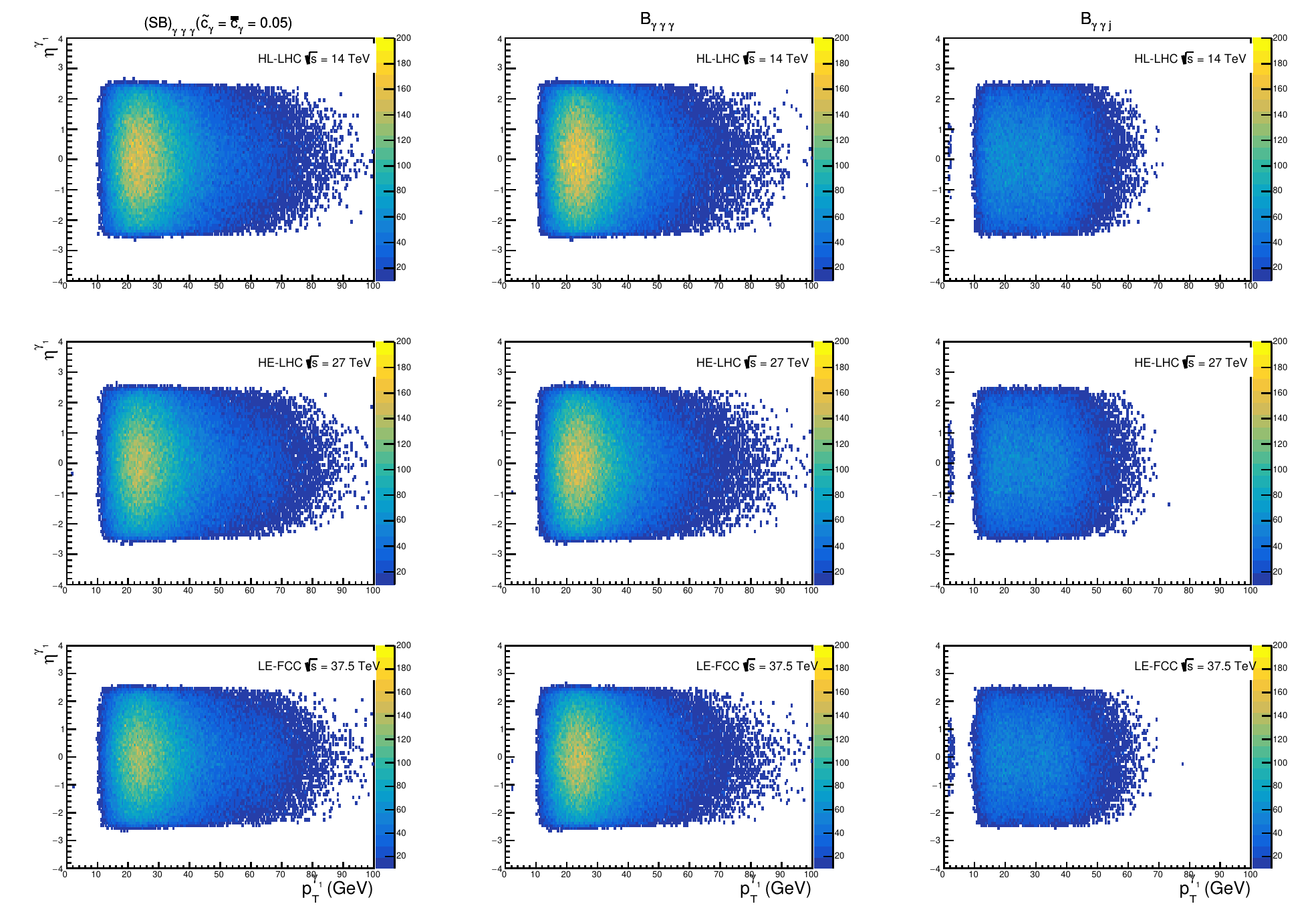} 
\caption{The phase space of the leading photon for signal ($SB_{\gamma\gamma\gamma}$) including SM and their interference, SM Background process ($B_{SM}$) with the same final state as signal and $B_{\gamma\gamma j}$ in each column. Rows are for different post-LHC circular collider.
\label{gamma1}}
\end{figure}

\begin{figure}[htb!]
\includegraphics[scale=0.8]{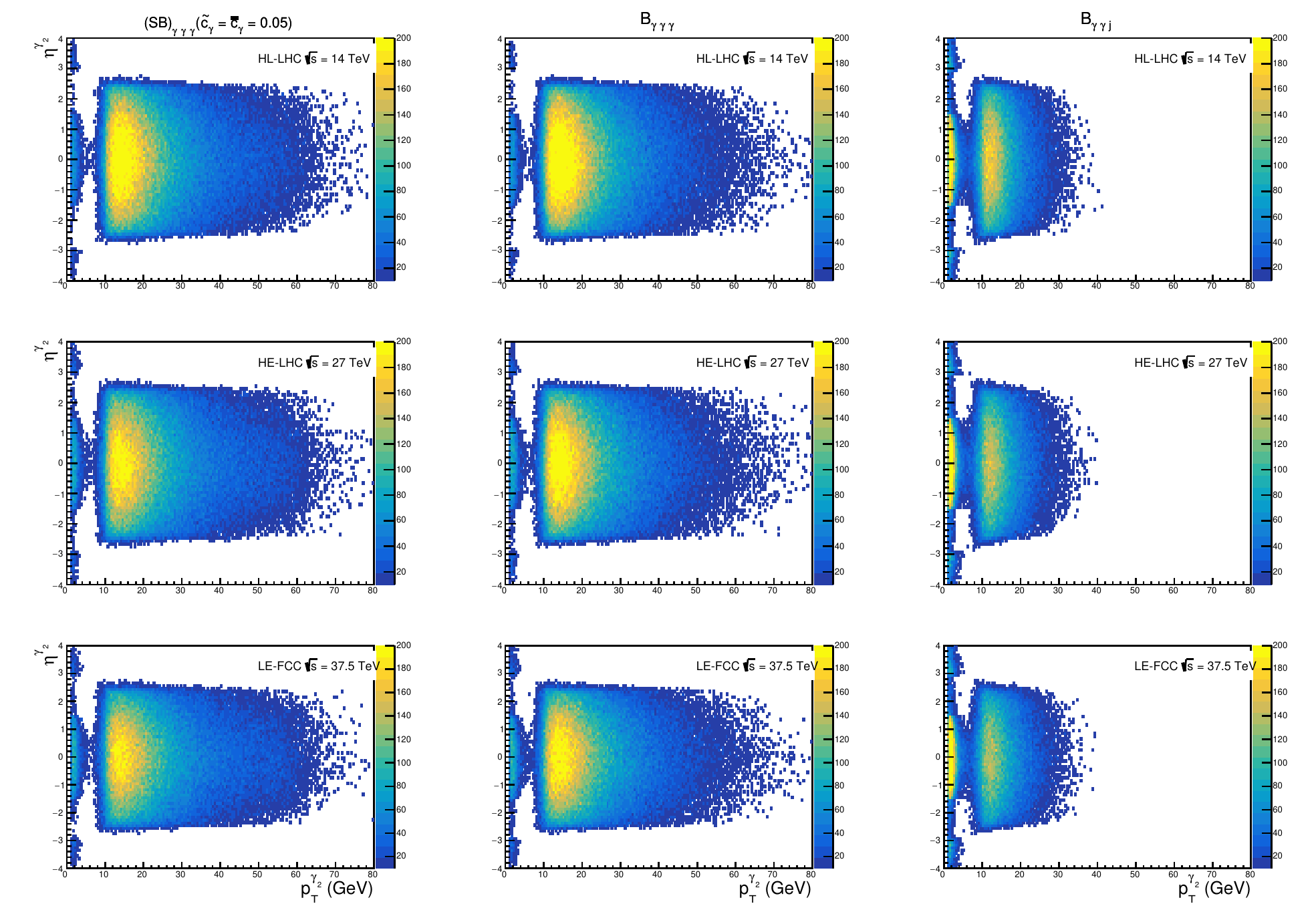} 
\caption{The phase space of the second-leading photon for signal ($SB_{\gamma\gamma\gamma}$) including SM and their interference, SM Background process ($B_{SM}$) with the same final state as signal and $B_{\gamma\gamma j}$ in each column. Rows are for different post-LHC circular collider.
\label{gamma2}}
\end{figure}

\begin{figure}[htb!]
\includegraphics[scale=0.8]{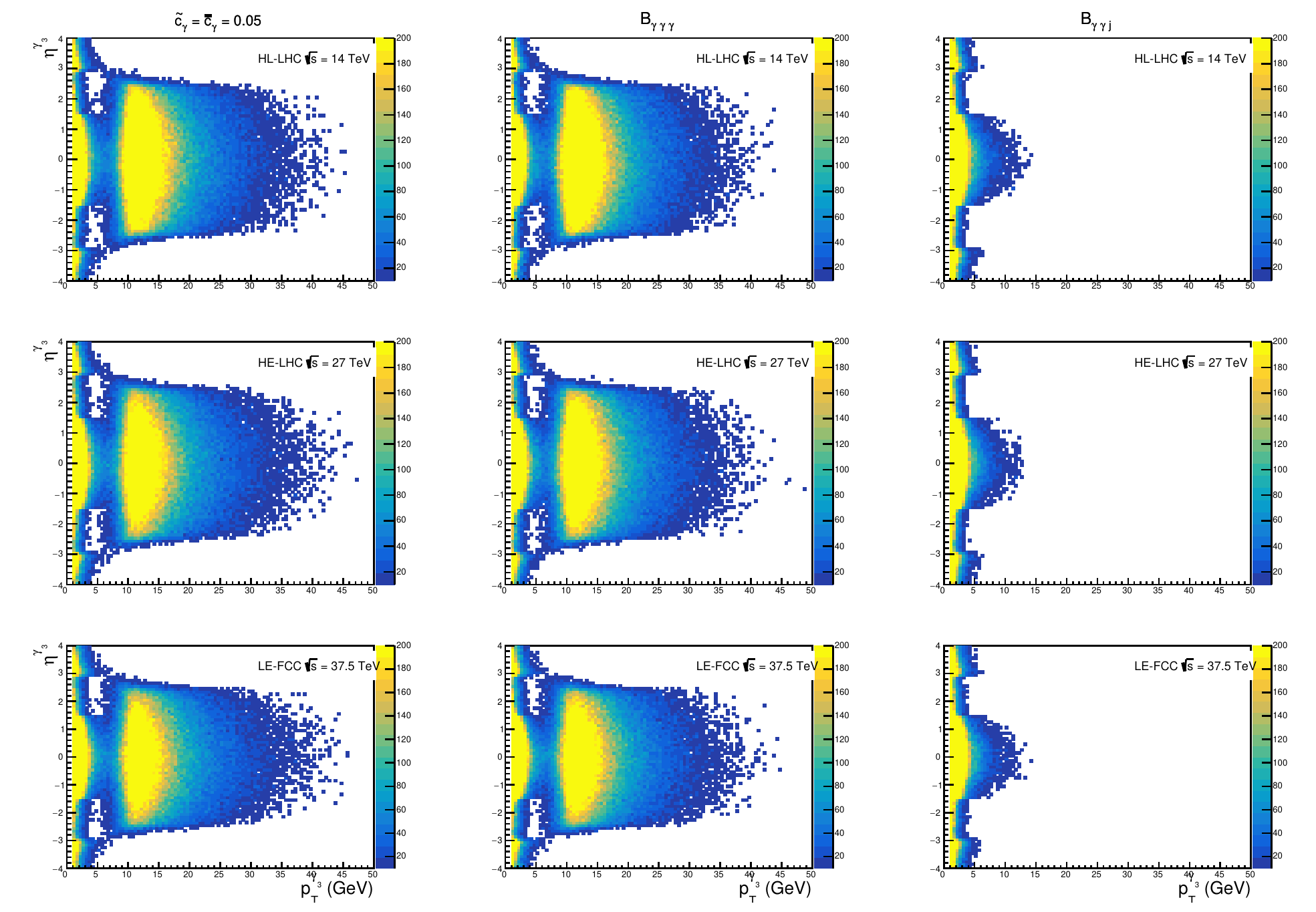} 
\caption{The phase space of the third-leading photon for signal ($SB_{\gamma\gamma\gamma}$) including SM and their interference, SM Background process ($B_{\gamma\gamma\gamma}$) with the same final state as signal and $B_{\gamma\gamma j}$ in each column. Rows are for different post-LHC circular collider.
\label{gamma3}}
\end{figure}

\begin{figure}[htb!]
\includegraphics[scale=0.8]{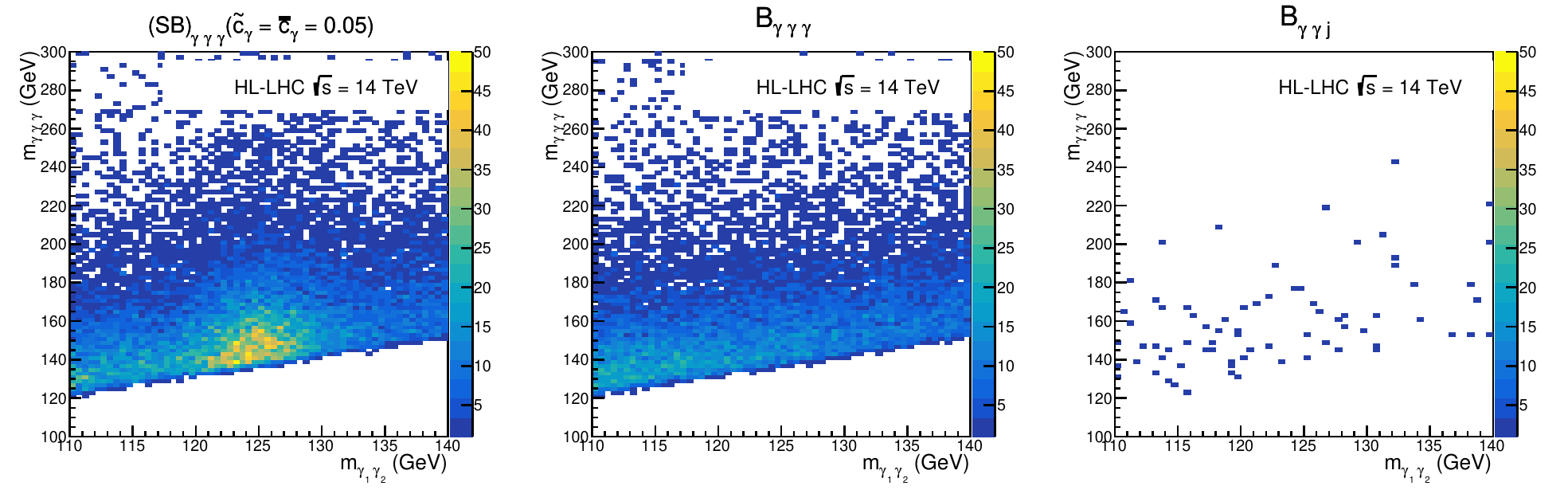} 
\caption{Invariant mass of three-photon versus the invariant mass of two-photon distribution for signal ($\bar{c}_{\gamma} =\tilde{c}_{\gamma} $=0.05) and 
relevant SM Backgrounds after Cut-4 for 14 TeV center of mass energy collider, namely HL-LHC (left-to-right).
\label{maaa_HL}}
\end{figure}

\begin{figure}[htb!]
\includegraphics[scale=0.8]{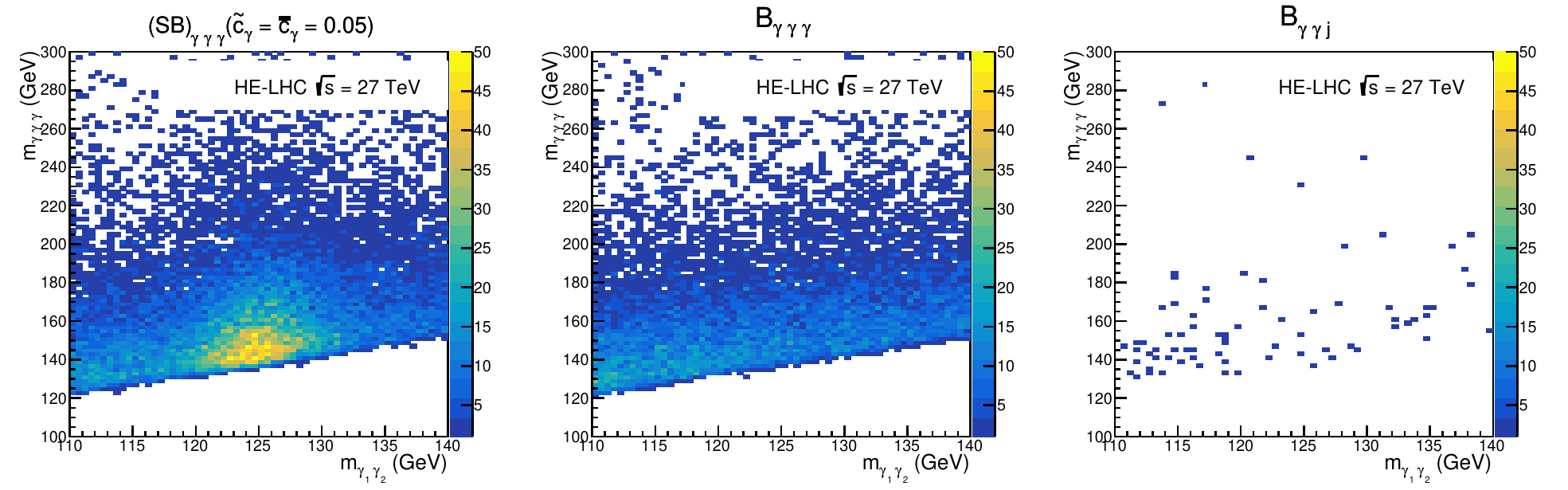} 
\caption{Invariant mass of three-photon versus the invariant mass of two-photon distribution for signal ($\bar{c}_{\gamma} =\tilde{c}_{\gamma} $=0.05) and 
relevant SM Backgrounds after Cut-4 for 27 TeV center of mass energy collider, namely HE-LHC.
\label{maaa_HE}}
\end{figure}

\begin{figure}[htb!]
\includegraphics[scale=0.8]{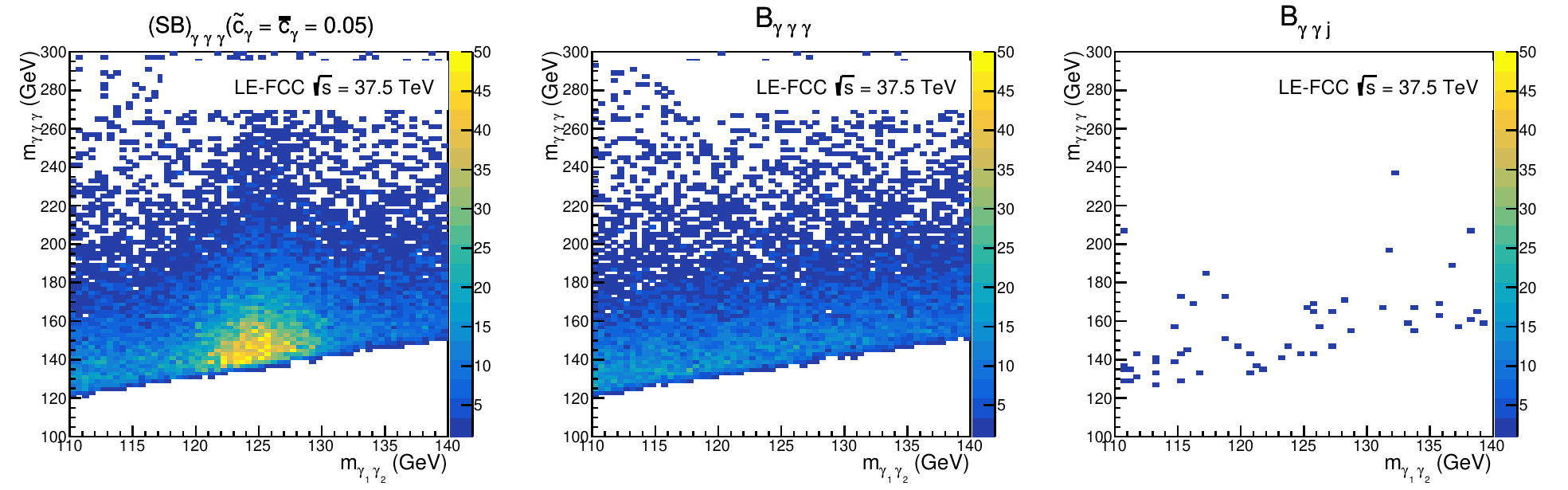} 
\caption{Invariant mass of three-photon versus the invariant mass of two-photon distribution for signal ($\bar{c}_{\gamma} =\tilde{c}_{\gamma} $=0.05) and 
relevant SM Backgrounds after Cut-4 for 37.5 TeV center of mass energy collider, namely LE-FCC.
\label{maaa_LE}}
\end{figure}

\begin{figure}[htb!]
\includegraphics[scale=0.27]{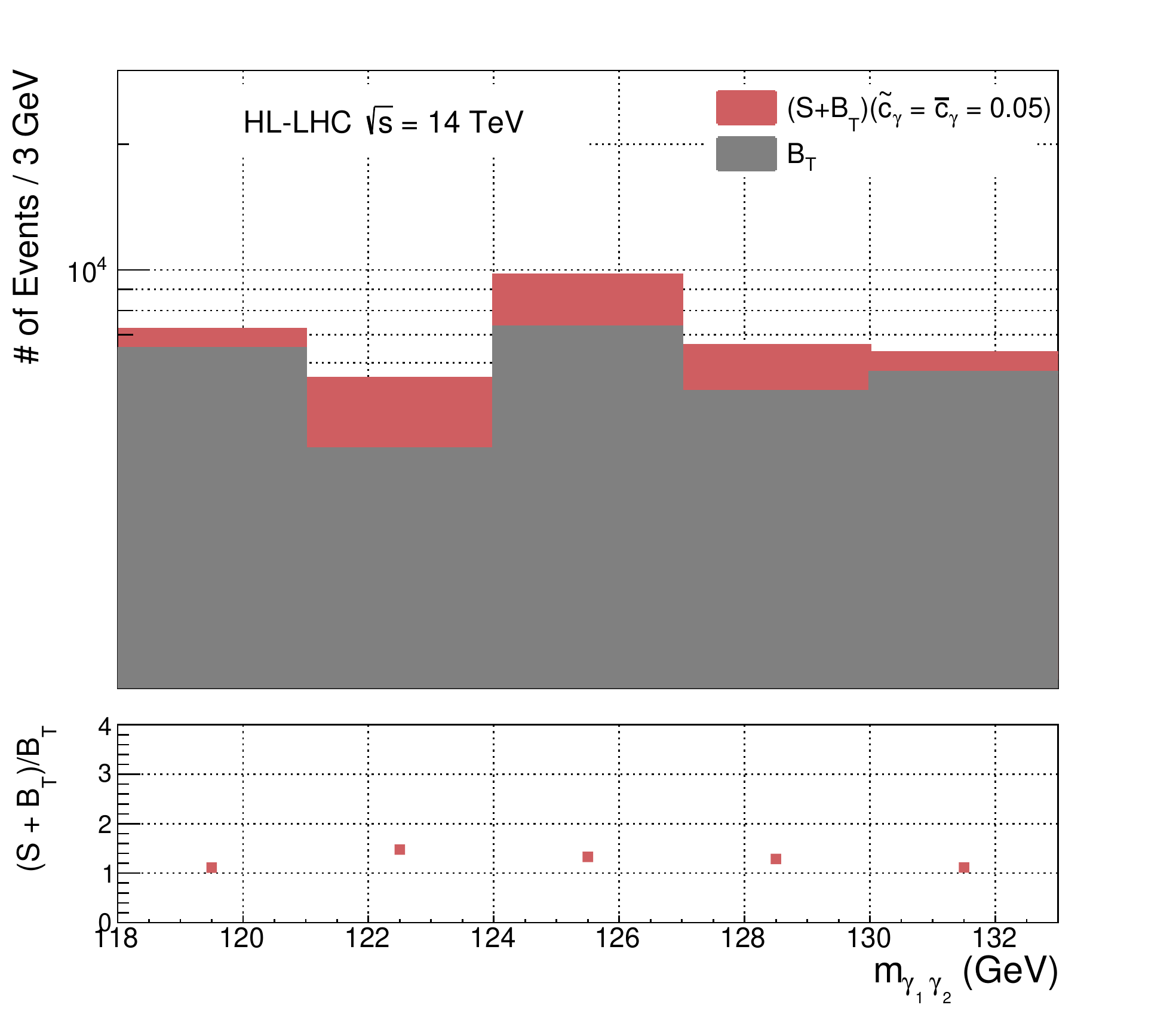} 
\includegraphics[scale=0.27]{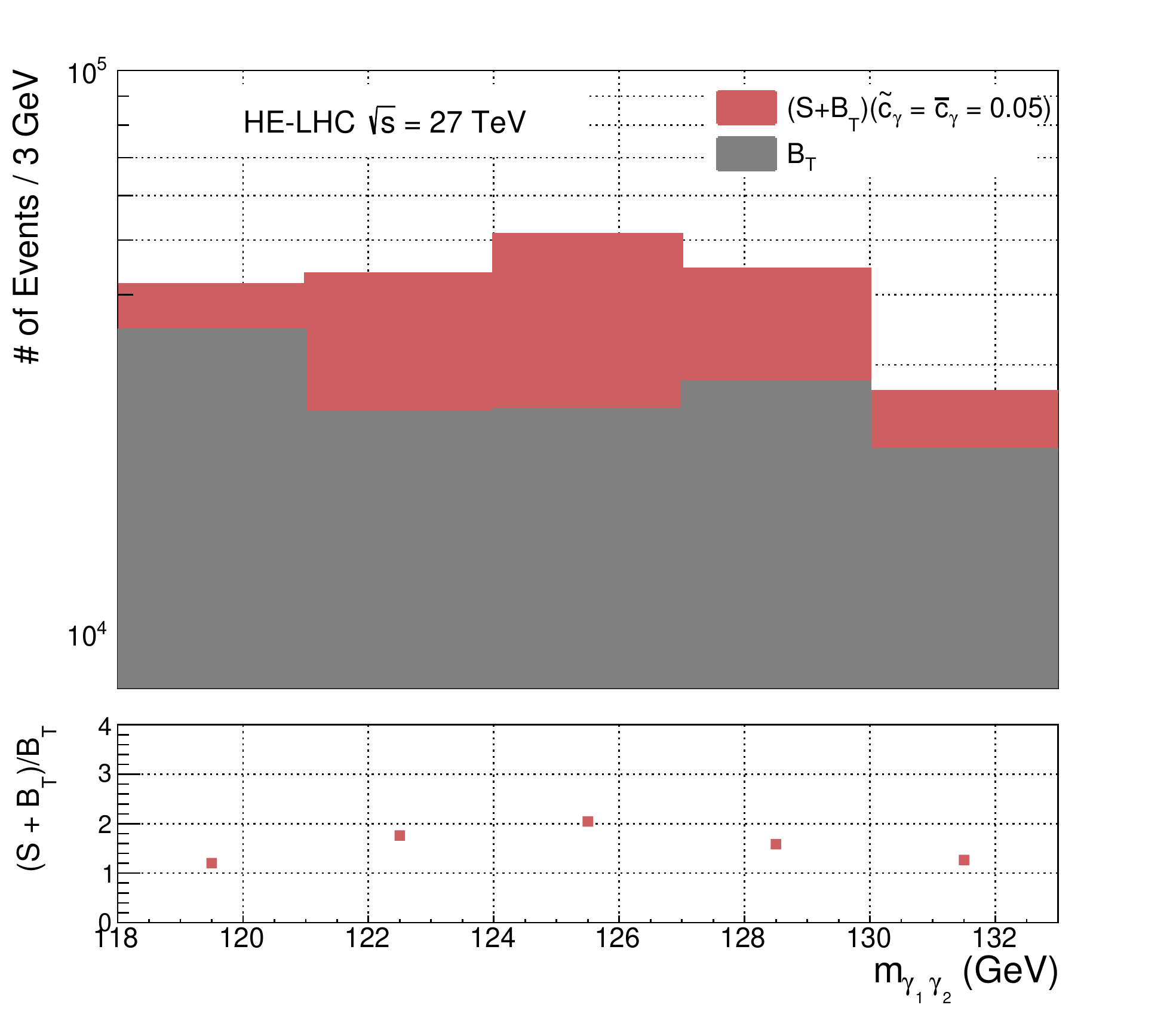} 
\includegraphics[scale=0.27]{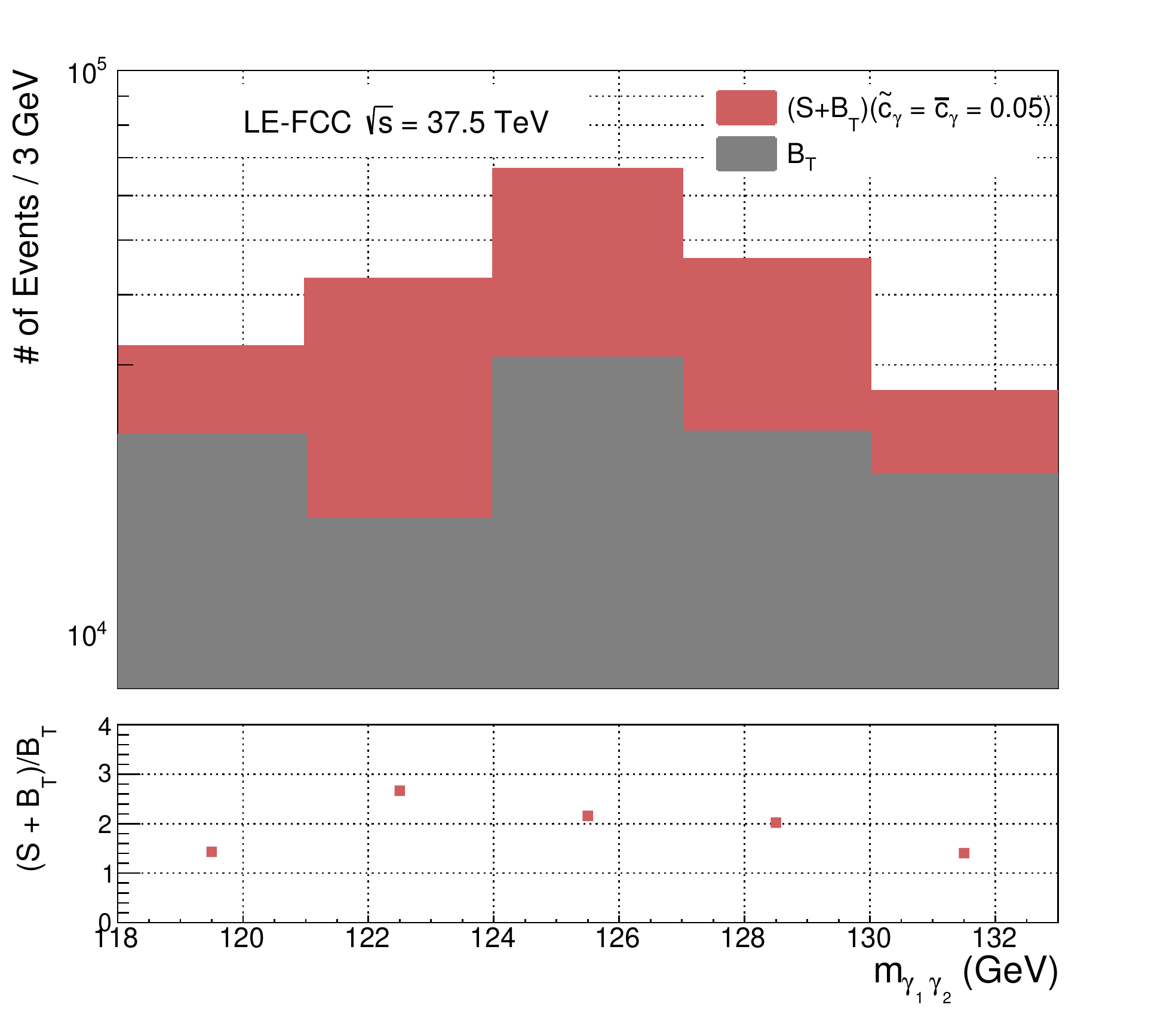} 
\caption{Invariant mass distribution of two-photon after cut-5 for $S+B_T$ ($\bar{c}_{\gamma} =\tilde{c}_{\gamma} $=0.05) (red) and total SM Background $B_T$ (gray) for HL-LHC, HE-LHC and LE-FCC, respectively. These distributions are normalized to relevant $L_{int}$ ( 3 ab$^{-1}$, 15 ab$^{-1}$ and 15 ab$^{-1}$).
\label{maa_all}}
\end{figure}

Even though all three collider configuration expect to see pile-up effects in the range of several hundreds, we did not consider any pile-up effects since we aim to give an estimation to obtain the limits of effective Higgs couplings by post-LHC circular high-energy hadron collider options through the production of three photons in this study \cite{Mangano:2020sao}. We filtered events with at least three photons in the final state and jet veto to suppress jet-containing backgrounds for the analysis as a first step (Cut-1). Photons are ordered with respect to their transverse momentum. That is, $\gamma_1$, $\gamma_2$ and $\gamma_3$ are the first, second and third leading photon, respectively ($p_T^{\gamma_1} > p_T^{\gamma_2} > p_T^{\gamma_3}$). Then, we review various kinematic variables of photons in order to use in “cut-based” analysis and to achieve physical intuition. The phase space of the first, second and third leading photons for signal with values $\bar{c}_{\gamma} $=0.05 and $\tilde{c}_{\gamma} $=0.05 and relevant SM backgrounds for HL-LHC, HE-LHC and LE-FCC are shown in Fig.~\ref{gamma1}, Fig.~\ref{gamma2} and Fig.~\ref{gamma3}, respectively. 
 These phase space distributions led us to the $p_T^{{\gamma_1}}>50$ GeV, $p_T^{{\gamma_2}}>35$ GeV, $p_T^{{\gamma_3}}>12$ GeV 
and $|\eta^{\gamma_{1,2,3}}| < 2.5$ region where the signal process can be separated from the backgrounds (Cut-2 and Cut-3). 
The distance between each photon is determined as $\Delta R(\gamma_i,\gamma_j)= \left[(\Delta\phi_{\gamma_i,\gamma_j}])^2+(\Delta\eta_{\gamma_i,\gamma_j}])^2\right]^{1/2}$ where $\Delta\phi_{\gamma_i,\gamma_j}$  and $\Delta\eta_{\gamma_i,\gamma_j}$ are the azimuthal angle and the pseudo rapidity difference between any two photons, respectively. Useful requirement to select isolated photons is to apply to the minimum distance between each photon as $\Delta R(\gamma_1,\gamma_2) > 0.4$, $\Delta R(\gamma_1,\gamma_3) > 0.4$, $\Delta R(\gamma_2,\gamma_3) > 0.4$ (Cut-4). Having the targeted signature with three prompt photons, we consider invariant mass of three photon as an important kinematic variable to extract limits on the couplings. Therefore, the distributions of invariant mass of three-photon versus the invariant mass of two-photon are checked for signal and relevant SM Backgrounds after Cut-4 at HL-LHC as well as other post-LHC hadron-hadron colliders under consideration. The distributions for signal with $\bar{c}_{\gamma} =\tilde{c}_{\gamma} $=0.05 couplings and relevant SM Backgrounds after Cut-4 (left-to-right) are given in Fig.~\ref{maaa_HL}, Fig.~\ref{maaa_HE} and Fig.~\ref{maaa_LE} for HL-LHC, HE-LHC and LE-FCC, respectively. We select events with invariant mass of three-photon  $m_{\gamma_1\gamma_2\gamma_3} > 160$ GeV (Cut-5). In order to focus on events where two photons are coming from decay of Higgs boson, we consider reconstructed invariant mass from two leading photons in the range of 122 GeV $< m_{\gamma\gamma}< 128 $ GeV (Cut 6). A summary of the cuts used in the analysis is given in Table \ref{cuts1}. 
\begin{table}[htb!]
\caption{List of optimized cuts considered in the analysis for selecting events to obtain limits. \label{cuts1}}
\begin{ruledtabular}
\begin{tabular}{llc}
 Cuts &Definations\\ \hline
Cut-1 &  $N_\gamma > 2 $, $N_{jet}=0$  \\
Cut-2&Cut-1+ $p_T^{{\gamma_1}}>50$ GeV, $p_T^{{\gamma_2}}>35$ GeV, $p_T^{{\gamma_3}}>12$ GeV \\
Cut-3& Cut-2+$|\eta^{\gamma_{1,2,3}}| < 2.5$\\
Cut-4& Cut-3+$\Delta R(\gamma_1,\gamma_2) > 0.4$, $\Delta R(\gamma_1,\gamma_3) > 0.4$, $\Delta R(\gamma_2,\gamma_3) > 0.4$ \\
Cut-5& Cut-4+$m_{\gamma_1\gamma_2\gamma_3} > 160$ GeV\\
Cut-6& Cut-5+122 GeV $< m_{\gamma\gamma}< 128 $ GeV\\
\end{tabular}
\end{ruledtabular}
\end{table}
The distributions of the reconstructed invariant mass of two leading photons is presented for the signal plus total SM backgrounds $S+B_T$ ($\bar{c}_{\gamma} =\tilde{c}_{\gamma} $=0.05) (red) and total SM Background $B_T$= $B_{\gamma\gamma\gamma}$+$B_{\gamma\gamma j}$ (gray) as well as their ratio ($S+B_T$)/$B_T$ in Fig.\ref{maa_all} for HL-LHC, HE-LHC and LE-FCC (left to right). 
Here, the main contribution comes from the $B_{\gamma\gamma\gamma}$ background. Number of events after this final cut is used to obtain limits on the anomalous Higgs effective couplings. Number of signal ($\bar{c}_{\gamma} $=$\tilde{c}_{\gamma} $=0.05) and relevant background events normalized to the corresponding luminosities 3 ab$^{-1}$, 15 ab$^{-1}$ and 15 ab$^{-1}$ (HL-LHC, HE-LHC and LE-FCC, respectively) after each cut given in Table~\ref{cuts}. Efficiency of each cut steps can be calculated from this table. Overall effect of the cuts used in the Table~\ref{cuts} changes between 0.8\% and  1.1\%  moving from HL-LHC  to LE-FCC for $\bar{c}_{\gamma} =\tilde{c}_{\gamma} $=0.05. On the other hand, efficiency of cuts for SM backgrounds $B_{\gamma\gamma\gamma}$ and  $B_{\gamma\gamma j}$ are 0.34\% and 0.003\%, respectively. Efficiency of the cuts also depend on the anomalous Higgs boson dimension-6 couplings value. We observed that efficiency gets lower to the 0.4\% for the signal with couplings set to $\bar{c}_{\gamma} =\tilde{c}_{\gamma} $=0.01. One might get better limits when cuts are optimized to each collider option. 
\begin{table}[htb!]
\caption{Number of signal and relevant events after each cut used in the analysis with integrated luminosities of 3 ab$^{-1}$, 15 ab$^{-1}$ and 15 ab$^{-1}$ for HL-LHC, HE-LHC and LE-FCC, respectively. \label{cuts}}
\begin{ruledtabular}
\begin{tabular}{lccccccc}
  Colliders&Process&Cut-1&Cut-2&Cut-3&Cut-4&Cut-5&Cut-6\\  \hline
& SB$_{\gamma\gamma\gamma}(\bar{c}_{\gamma}=\tilde{c}_{\gamma} =0.05)$&    373073 & 54673.7 & 53502.3 & 53275.4  &38071.9  &2916.41\\ 
  HL-LHC& $B_{\gamma\gamma\gamma}$ &278456  &36454.8&  35640.1  &35491.9 & 26282.3  &954.18\\ 
   &$B_{\gamma\gamma j}$  &1.2704 x$10^{8}$&   158052&   144823 &  139949 &  78677.8  &3481.32\\ \hline
 & SB$_{\gamma\gamma\gamma}(\bar{c}_{\gamma}=\tilde{c}_{\gamma} =0.05)$  &2.91392x$10^{6}$ & 450225  & 440311  & 437960 &  310376  & 29650.2  \\ 
  HE-LHC &$B_{\gamma\gamma\gamma}$ &2.13519x$10^{6}$ &  283158  & 276893 &   275816 &  205528 &  7306.28 \\ 
 & $B_{\gamma\gamma j}$ &5.98137x$10^{8}$&  710189 &   644044 &  637082 &   337688 &  20887.9 \\ \hline
 &SB$_{\gamma\gamma\gamma}(\bar{c}_{\gamma}=\tilde{c}_{\gamma} =0.05)$  &3.61827x$10^{6}$&  573519 &  559763  & 556825 &  395603  &41175 \\ 
  LE-FCC& $B_{\gamma\gamma\gamma}$ &2.63292x$10^{6}$&  349711 & 341548  & 340030 &  254747 & 9522.67  \\ 
& $B_{\gamma\gamma j}$&5.70719x$10^{8}$&  696264 &  647526&   633600   &382945  & 13925.3   \\ 
\end{tabular}
\end{ruledtabular}
\end{table}

\section{Sensitivity of the dimension-6 Higgs-gauge boson couplings }
\begin{figure}[htb!]
\includegraphics[scale=0.45]{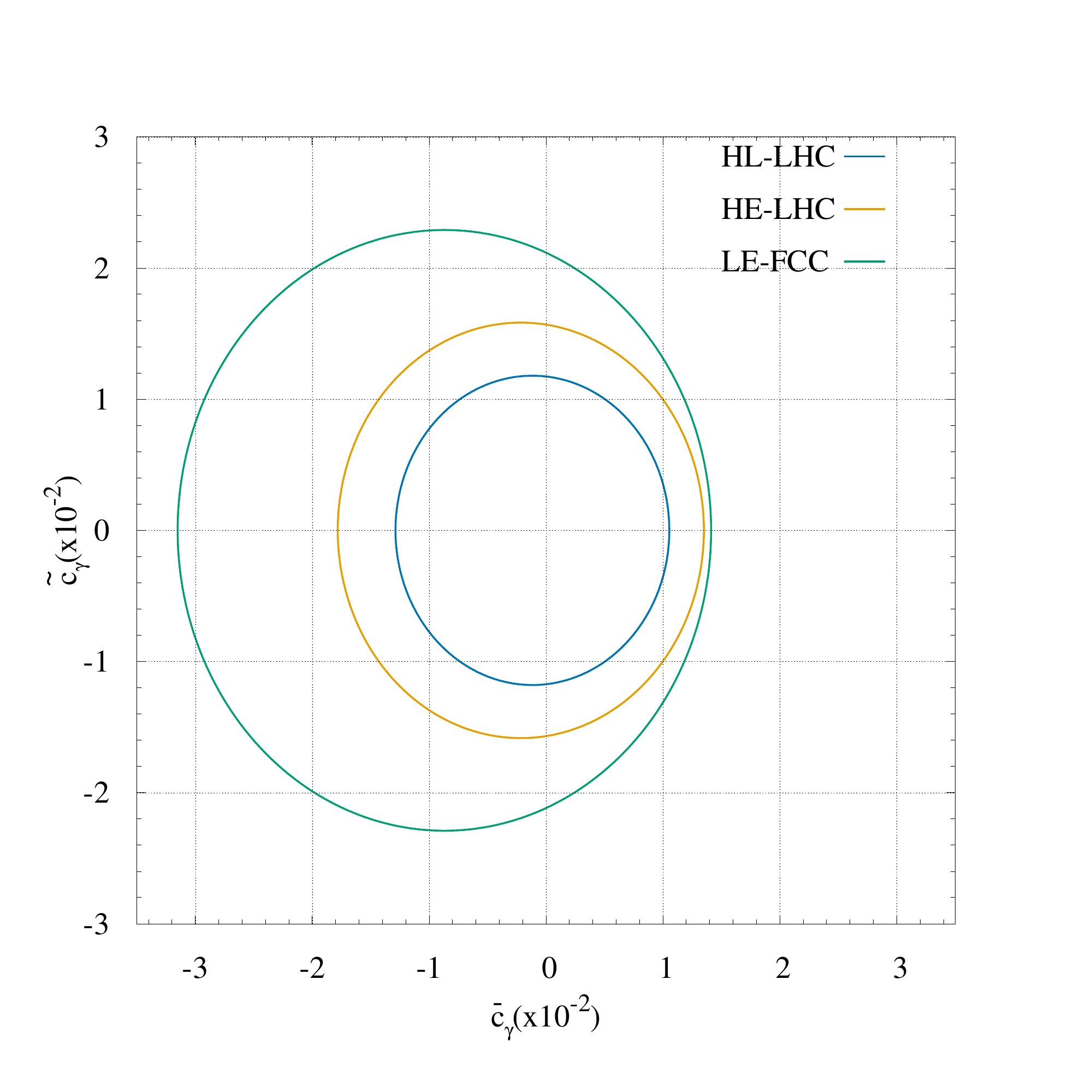} 
\caption{Two-dimensional 95\% C.L. intervals in plane for $\bar{c}_{\gamma}$ and $\tilde{c}_{\gamma}$ without of systematic errors for HL-LHC, HE-LHC and LE-FCC taking $L_{int}= 3, 15$ and 15  ab$^{-1}$ respectively. The limits are  derived with all other coefficients set to zero.
\label{limits_all}}
\end{figure}

\begin{figure}[htb!]
\includegraphics[scale=0.26]{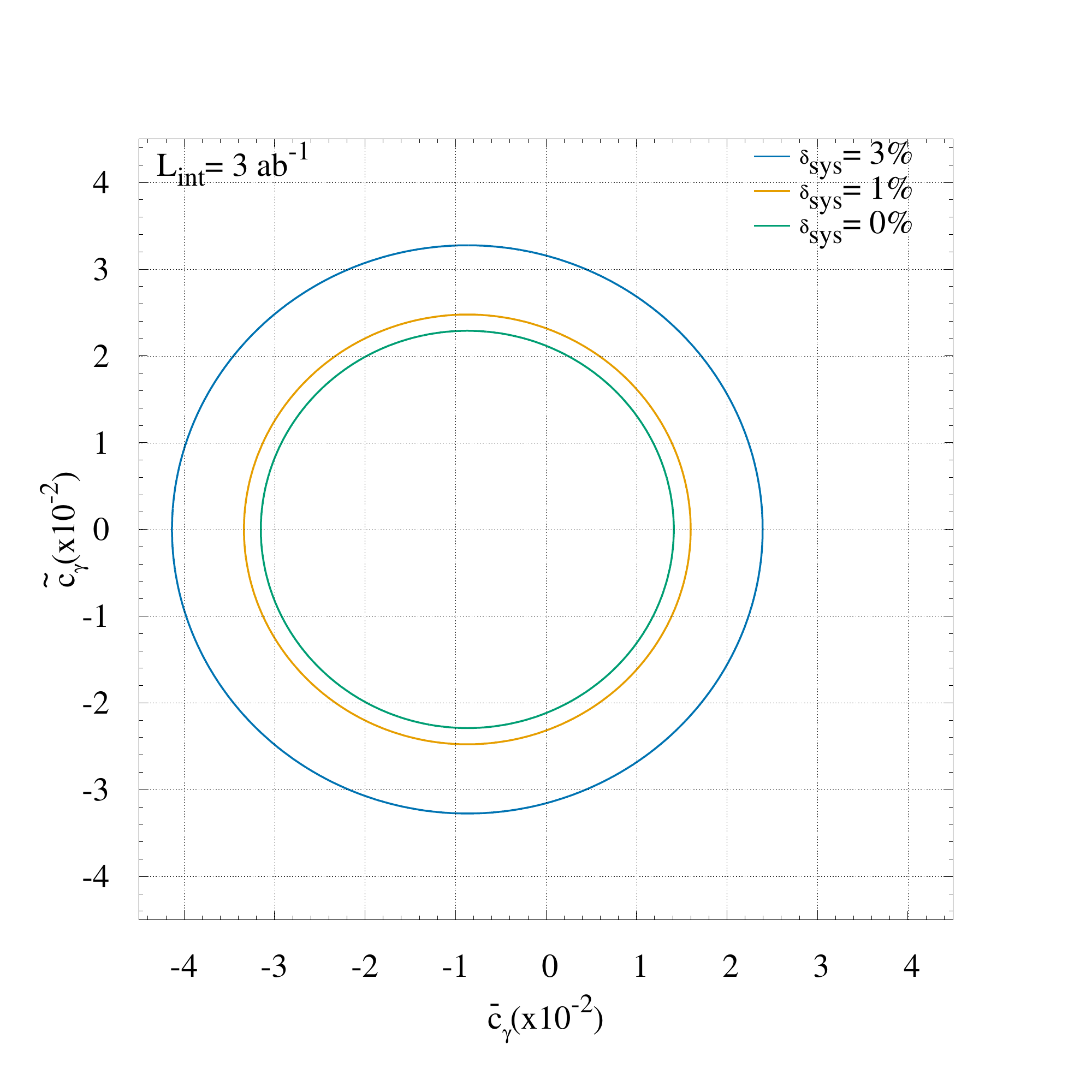} 
\includegraphics[scale=0.26]{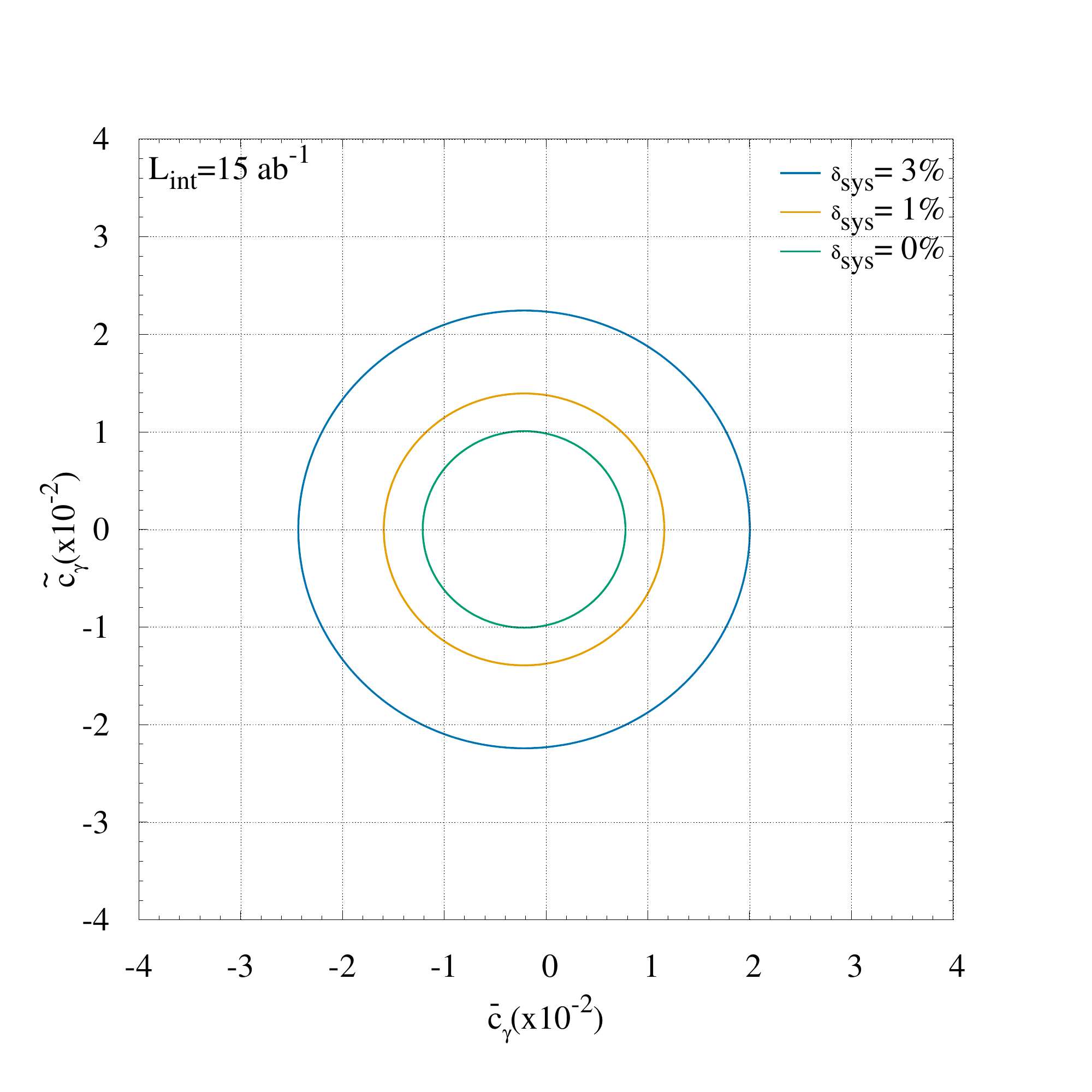} 
\includegraphics[scale=0.26]{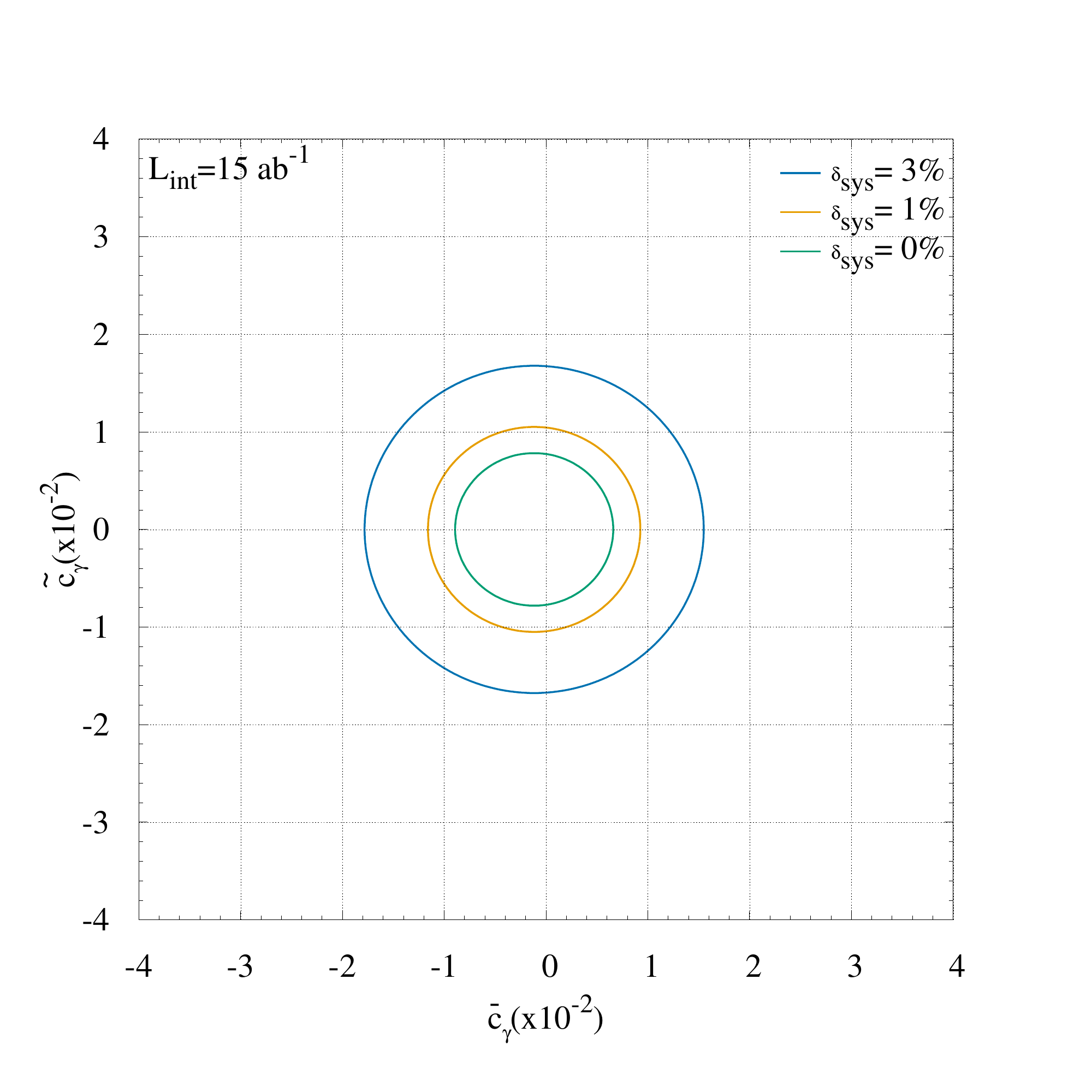} 
\caption{Two-dimensional 95\% C.L. intervals on the $\bar{c}_{\gamma}$ and $\tilde{c}_{\gamma}$ couplings plane considering 0, 1 \% and 3\% systematic error at HL-LHC, HE-LHC and LE-FCC (left-to-right). The limits are derived with all coefficients other than $\bar{c}_{\gamma}$ and $\tilde{c}_{\gamma}$ set to zero.
\label{limits}}
\end{figure}

The $\chi^{2}$ statistical analysis approach, which measures how the expectations are compared with the actual data observed (or model results), is used to obtain the sensitivity of the dimension-6 Higgs-gauge boson couplings in $pp\to\gamma\gamma\gamma$ process as follows;
\begin{eqnarray}
\chi^{2} =\sum_i^{n_{bins}}\left(\frac{N_{i}^{NP}-N_{i}^{B}}{N_{i}^{B}\Delta_i}\right)^{2}
\end{eqnarray}
where $N_i^{NP}$ is the total number of events in the existence of effective couplings ($S$) , the number of events of relevant SM backgrounds in $i$th bin of the invariant mass distributions of reconstructed Higgs boson from two leading photon denotes $N_i^B$ , $\Delta_i=\sqrt{\delta_{sys}^2+\frac{1}{N_i^B}}$ is the combined systematic ($\delta_{sys}$) and statistical errors in each bin. In this analysis, we focused on $\bar{c}_{\gamma} $ and $\tilde{c}_{\gamma} $ couplings which are the main coefficients contributing to $pp\to\gamma\gamma\gamma$ signal process. In two-dimensional $\chi^{2}$ analysis, two Higgs-gauge boson couplings $\bar{c}_{\gamma}$ and $\tilde{c}_{\gamma}$ are assumed to deviate from their SM values simultaneously while all other Wilson coefficients
set to zero. In Fig. \ref{limits_all}, we show 95\% C.L. contours for anomalous $\bar{c}_{\gamma}$ and $\tilde{c}_{\gamma}$ couplings with integrated luminosities of 3 ab$^{-1}$, 15 ab$^{-1}$ and 15 ab$^{-1}$ for HL-LHC, HE-LHC and LE-FCC without systematic errors. As we can see from Fig.\ref{limits_all}, the best limits without systematic error on dimension-6 Higgs-gauge boson couplings $\bar{c}_{\gamma}$ ( $\tilde{c}_{\gamma}$) couplings are $[-3.15;1.41]\times10^{-2}$ ($[-2.12;2.12]\times10^{-2}$), $[-1.21;0.78]\times10^{-2}$ ($[-0.98;0.98]\times10^{-2}$) and $[-0.89;0.66]\times10^{-2}$ ($[-0.77;0.77]\times10^{-2}$) for HL-LHC, HE-LHC and LE-FCC, respectively.  In Fig. \ref{limits}, we also present same contour plot taking into account systematic error for HL-LHC, HE-LHC and LE-FCC, respectively. The limits on $\bar{c}_{\gamma}$ and $\tilde{c}_{\gamma}$ couplings get worse when systematic errors are increased in each hadron colliders considered in this study. For example, when $\delta_{sys}$ =\%3, the limits on $\bar{c}_{\gamma}$ ( $\tilde{c}_{\gamma}$) couplings is $[-1.67;1.67]\times10^{-2}$ ($[-1.69;1.69]\times10^{-2}$) for LE-FCC collider with 15 ab$^{-1}$ of integrated luminosity. These limits are up to 2 times worse than those obtained without systematic errors as seen from Fig. \ref{limits}. ATLAS experiment probed limits on these couplings by using a fit to five measured differential cross sections in $H\to\gamma\gamma$ decay channel \cite{Aad:2015tna}. They obtained [-7.4; 5.7]$\times10^{-4}$ $\cup$ [3.8; 5.1]$\times10^{-3}$ and [-1.8; 1.8]$\times10^{-3}$ limits on $\bar{c}_{\gamma} $ and $\tilde{c}_{\gamma}$, respectively with an integrated luminosity of 20.3 fb$^{-1}$ at $\sqrt s$=8 TeV, respectively. In their similar analysis on 13 TeV center of mass energy with an integrated luminosity of 36.1 fb$^{-1}$, they claim that  $H\to\gamma\gamma$ decay channel is not sensitive to $\bar{c}_{\gamma}$ and $\tilde{c}_{\gamma}$ \cite{Aaboud:2018xdt}. Results of the analysis with increased luminosity ($L_{int}$=139 fb$^{-1}$) at $\sqrt s$=13 TeV by ATLAS collaboration are [-1.1; 1.1]$\times10^{-4}$  and [-2.8; 4.3]$\times10^{-4}$  for $\bar{c}_{\gamma} $ and $\tilde{c}_{\gamma}$, respectively \cite{ATLAS:2019jst}. In our study, limitations are placed on dimension-6 operators by focusing only on three-photon production in the post-LHC scenarios. One can obtain stringent bounds on these operators by taking into account for other decay channels of the Higgs boson as in Refs. \cite{Aad:2015tna, Aaboud:2018xdt, ATLAS:2019jst}. Furthermore, it is also possible to include three-photon analysis in global fit with other production processes to achieve limits on dimension-6 operators with greater precision.

\section{Conclusions}
The potential of $pp\to \gamma\gamma\gamma$ process is investigated to obtain limits on the $\bar{c}_{\gamma}$ and $\tilde{c}_{\gamma}$ couplings at 95\% confidence level at High-Luminosity LHC (HL-LHC) as well as other post-LHC hadron-hadron colliders under consideration (14 TeV/3 ab$^{-1}$, 27 TeV/15 ab$^{-1}$ and 37.5 TeV/15 ab$^{-1}$, respectively). Signal (non-zero couplings) including interference with SM and both background events are generated in MadGraph where the effective Lagrangian of the SM EFT is implemented using FeynRules and UFO framework. Then, events are passed through PYTHIA for parton showering and hadronization and Delphes to include realistic detector effects. 64 samples for each hadron collider consideration are generated by varying two Wilson coefficients simultaneously to obtain sensitivity bounds on the couplings. The targeted signature consists of three prompt photons. Therefore, 2D plots of kinematic variables pseudo-rapidity versus transverse momentum of each photon and invariant mass distributions of three photon as function of reconstructed invariant mass of two leading photons are plotted to determine a cut-based analysis.   To identify the signal over background, we made a series of standard cuts on the transverse momentum and pseudorapidity of three leading photon as well as photon separation $\Delta R$. We also apply cut on invariant mass of two-photon system reconstructed form two leading photons. Finally we use transverse momentum of three-photon system for $\chi^2$ analysis to obtain limits. The reconstructed invariant mass of three photon in the range of Higgs-boson reconstructed from two leading photons is used to obtain limits on the anomalous Higgs effective couplings in $pp\to \gamma\gamma\gamma$ signal process and the relevant SM background. Our obtained limits without systematic error on $\bar{c}_{\gamma}$ ( $\tilde{c}_{\gamma}$) are  $[-3.15;1.41]\times10^{-2}$ ($[-2.12;2.12]\times10^{-2}$), $[-1.21;0.78]\times10^{-2}$ ($[-0.98;0.98]\times10^{-2}$) and $[-0.89;0.66]\times10^{-2}$ ($[-0.77;0.77]\times10^{-2}$) for HL-LHC, HE-LHC, LE-FCC, respectively. No discussion of possible sources of systematic uncertainty is given in the manuscript. But its effects on the limits of the couplings are considered. Results including systematic errors get worse as expected. However, we predict the testable bounds from these post-LHC colliders via tri-photon production on the anomalous Higgs boson couplings even with 1\% systematic uncertainty from possible experimental sources.

\begin{acknowledgments}
This work was partially supported by Turkish Atomic Energy Authority (TAEK) under the grant No. 2018TAEK(CERN)A5.H6.F2-20. \end{acknowledgments}


\begin{thebibliography}{99}
\bibitem{Aad:2012tfa} 
  G.~Aad {\it et al.} [ATLAS Collaboration],
  Phys.\ Lett.\ B {\bf 716}, 1 (2012)
  [arXiv:1207.7214 [hep-ex]].
  
\bibitem{Chatrchyan:2012xdj} 
  S.~Chatrchyan {\it et al.} [CMS Collaboration],
  Phys.\ Lett.\ B {\bf 716}, 30 (2012)
  [arXiv:1207.7235 [hep-ex]].

\bibitem{ApollinariG.:2017ojx} 
  G.~Apollinari, I.~Béjar Alonso, O.~Brüning, P.~Fessia, M.~Lamont, L.~Rossi and L.~Tavian,
  CERN Yellow Rep.\ Monogr.\  {\bf 4}, 1 (2017).

\bibitem{Abada:2019ono} 
  A.~Abada {\it et al.} [FCC Collaboration],
  Eur.\ Phys.\ J.\ ST {\bf 228}, no. 5, 1109 (2019).
  
\bibitem{Mangano:2681366}
M.~Mangano,
CERN-FCC-PHYS-2019-0001, "https://cds.cern.ch/record/2681366".
  
 \bibitem{Buchmuller:1985jz} 
  W.~Buchmuller and D.~Wyler,
  Nucl.\ Phys.\ B {\bf 268}, 621 (1986).
  
  
  
\bibitem{Hagiwara:1993qt} 
  K.~Hagiwara, R.~Szalapski and D.~Zeppenfeld,
  Phys.\ Lett.\ B {\bf 318}, 155 (1993)
  [hep-ph/9308347].
  
\bibitem{Corbett:2012ja} 
  T.~Corbett, O.~J.~P.~Eboli, J.~Gonzalez-Fraile and M.~C.~Gonzalez-Garcia,
  Phys.\ Rev.\ D {\bf 87}, 015022 (2013)
  [arXiv:1211.4580 [hep-ph]].
  
\bibitem{Ellis:2014jta} 
  J.~Ellis, V.~Sanz and T.~You,
  JHEP {\bf 1503}, 157 (2015)
  [arXiv:1410.7703 [hep-ph]].
  
\bibitem{Ellis:2014dva} 
  J.~Ellis, V.~Sanz and T.~You,
  JHEP {\bf 1407}, 036 (2014)
  [arXiv:1404.3667 [hep-ph]].

\bibitem{Falkowski:2015fla} 
  A.~Falkowski,
  Pramana {\bf 87}, no. 3, 39 (2016)
  [arXiv:1505.00046 [hep-ph]].

\bibitem{Corbett:2015ksa} 
  T.~Corbett, O.~J.~P.~Eboli, D.~Goncalves, J.~Gonzalez-Fraile, T.~Plehn and M.~Rauch,
  JHEP {\bf 1508}, 156 (2015)
  [arXiv:1505.05516 [hep-ph]].
  
  \bibitem{Ferreira:2016jea} 
  F.~Ferreira, B.~Fuks, V.~Sanz and D.~Sengupta,
  Eur.\ Phys.\ J.\ C {\bf 77}, no. 10, 675 (2017)
  [arXiv:1612.01808 [hep-ph]].
  
\bibitem{Khachatryan:2014kca} 
  V.~Khachatryan {\it et al.} [CMS Collaboration],
  Phys.\ Rev.\ D {\bf 92}, no. 1, 012004 (2015)
  [arXiv:1411.3441 [hep-ex]].

\bibitem{Khachatryan:2016tnr} 
  V.~Khachatryan {\it et al.} [CMS Collaboration],
  Phys.\ Lett.\ B {\bf 759}, 672 (2016)
  [arXiv:1602.04305 [hep-ex]].
  
\bibitem{Khanpour:2017inb}
H.~Khanpour, S.~Khatibi and M.~Mohammadi Najafabadi,
Phys. Lett. B \textbf{773}, 462-469 (2017)
[arXiv:1702.05753 [hep-ph]].

\bibitem{vonBuddenbrock:2017gvy}
S.~von Buddenbrock, A.~S.~Cornell, A.~Fadol, M.~Kumar, B.~Mellado and X.~Ruan,
J. Phys. G \textbf{45}, no.11, 115003 (2018)
[arXiv:1711.07874 [hep-ph]].

\bibitem{Shi:2018lqf}
L.~Shi, Z.~Liang, B.~Liu and Z.~He,
Chin. Phys. C \textbf{43}, no.4, 043001 (2019)
[arXiv:1811.02261 [hep-ph]].

\bibitem{Freitas:2019hbk}
F.~F.~Freitas, C.~K.~Khosa and V.~Sanz,
Phys. Rev. D \textbf{100}, no.3, 035040 (2019)
[arXiv:1902.05803 [hep-ph]].

\bibitem{Aaboud:2019nan}
M.~Aaboud \textit{et al.} [ATLAS],
JHEP \textbf{05}, 141 (2019)
[arXiv:1903.04618 [hep-ex]].

\bibitem{Banerjee:2019jys}
S.~Banerjee, F.~Krauss and M.~Spannowsky,
Phys. Rev. D \textbf{100}, 073012 (2019)
[arXiv:1904.07886 [hep-ph]].
  
\bibitem{Aad:2015tna} 
  G.~Aad {\it et al.} [ATLAS Collaboration],
  Phys.\ Lett.\ B {\bf 753}, 69 (2016)
  [arXiv:1508.02507 [hep-ex]].
 
\bibitem{Englert:2015hrx} 
  C.~Englert, R.~Kogler, H.~Schulz and M.~Spannowsky,
  Eur.\ Phys.\ J.\ C {\bf 76}, no. 7, 393 (2016)
  [arXiv:1511.05170 [hep-ph]].

\bibitem{Buckley:2015lku} 
  A.~Buckley, C.~Englert, J.~Ferrando, D.~J.~Miller, L.~Moore, M.~Russell and C.~D.~White,
  JHEP {\bf 1604}, 015 (2016)
  [arXiv:1512.03360 [hep-ph]].
  
  \bibitem{Englert:2016hvy} 
  C.~Englert, R.~Rosenfeld, M.~Spannowsky and A.~Tonero,
  EPL {\bf 114}, no. 3, 31001 (2016)
  [arXiv:1603.05304 [hep-ph]].

\bibitem{Degrande:2016dqg} 
  C.~Degrande, B.~Fuks, K.~Mawatari, K.~Mimasu and V.~Sanz,
  Eur.\ Phys.\ J.\ C {\bf 77}, no. 4, 262 (2017)
  [arXiv:1609.04833 [hep-ph]].

\bibitem{Bishara:2016kjn} 
  F.~Bishara, R.~Contino and J.~Rojo,
  Eur.\ Phys.\ J.\ C {\bf 77}, no. 7, 481 (2017)
  [arXiv:1611.03860 [hep-ph]].
  
\bibitem{Liu-Sheng:2017pxk} 
  L.~S.~Ling, R.~Y.~Zhang, W.~G.~Ma, X.~Z.~Li, L.~Guo and S.~M.~Wang,
  Phys.\ Rev.\ D {\bf 96}, no. 5, 055006 (2017)
  [arXiv:1708.04785 [hep-ph]].

\bibitem{Aaboud:2018xdt} 
  M.~Aaboud {\it et al.} [ATLAS Collaboration],
  Phys.\ Rev.\ D {\bf 98}, 052005 (2018)
  [arXiv:1802.04146 [hep-ex]].
  
    

  
\bibitem{deCampos:1998xx} 
  F.~de Campos, M.~C.~Gonzalez-Garcia, S.~M.~Lietti, S.~F.~Novaes and R.~Rosenfeld,
  Phys.\ Lett.\ B {\bf 435}, 407 (1998)
  [hep-ph/9806307].
  
\bibitem{GonzalezGarcia:1999fq} 
  M.~C.~Gonzalez-Garcia,
  Int.\ J.\ Mod.\ Phys.\ A {\bf 14}, 3121 (1999)
  [hep-ph/9902321].

\bibitem{Aaboud:2017lxm} 
  M.~Aaboud {\it et al.} [ATLAS Collaboration],
  Phys.\ Lett.\ B {\bf 781}, 55 (2018)
  [arXiv:1712.07291 [hep-ex]].

\bibitem{Achard:2004kn} 
  P.~Achard {\it et al.} [L3 Collaboration],
  Phys.\ Lett.\ B {\bf 589}, 89 (2004)
  [hep-ex/0403037].

  \bibitem{Abreu:1999vt} 
  P.~Abreu {\it et al.} [DELPHI Collaboration],
  Phys.\ Lett.\ B {\bf 458}, 431 (1999).

\bibitem{Heister:2002ub} 
  A.~Heister {\it et al.} [ALEPH Collaboration],
  Phys.\ Lett.\ B {\bf 544}, 16 (2002).

\bibitem{Denizli:2019ijf}
H.~Denizli, K.~Oyulmaz and A.~Senol,
J. Phys. G \textbf{46} (2019) no.10, 105007
[arXiv:1901.04784 [hep-ph]].


\bibitem{Alloul:2013naa} 
  A.~Alloul, B.~Fuks and V.~Sanz,
  JHEP {\bf 1404}, 110 (2014)
  [arXiv:1310.5150 [hep-ph]].

\bibitem{Alwall:2014hca}
  J.~Alwall {\it et al.},
  JHEP {\bf 1407} (2014) 079
  [arXiv:1405.0301 [hep-ph]].

\bibitem{Alloul:2013bka} 
  A.~Alloul, N.~D.~Christensen, C.~Degrande, C.~Duhr and B.~Fuks,
  Comput.\ Phys.\ Commun.\  {\bf 185}, 2250 (2014)
 [arXiv:1310.1921 [hep-ph]].

\bibitem{Degrande:2011ua} 
  C.~Degrande, C.~Duhr, B.~Fuks, D.~Grellscheid, O.~Mattelaer and T.~Reiter,
  Comput.\ Phys.\ Commun.\  {\bf 183}, 1201 (2012)
  [arXiv:1108.2040 [hep-ph]].
%
\bibitem{Sjostrand:2014zea} 
  T.~Sjöstrand {\it et al.},
  Comput.\ Phys.\ Commun.\  {\bf 191}, 159 (2015)
  [arXiv:1410.3012 [hep-ph]].

  \bibitem{deFavereau:2013fsa} 
  J.~de Favereau {\it et al.} [DELPHES 3 Collaboration],
  JHEP {\bf 1402}, 057 (2014)
  [arXiv:1307.6346 [hep-ex]].
  
  \bibitem{Cacciari:2011ma} 
  M.~Cacciari, G.~P.~Salam and G.~Soyez,
  Eur.\ Phys.\ J.\ C {\bf 72}, 1896 (2012)
  [arXiv:1111.6097 [hep-ph]].

\bibitem{Cacciari:2008gp} 
  M.~Cacciari, G.~P.~Salam and G.~Soyez,
  JHEP {\bf 0804}, 063 (2008)
  [arXiv:0802.1189 [hep-ph]].
  
  \bibitem{exroot}
  http://madgraph.hep.uiuc.edu/Downloads/ExRootAnalysis

\bibitem{Brun:1997pa} 
  R.~Brun and F.~Rademakers,
  Nucl.\ Instrum.\ Meth.\ A {\bf 389}, 81 (1997).
  
\bibitem{Mangano:2020sao}
M.~L.~Mangano, G.~Ortona and M.~Selvaggi,
Eur. Phys. J. C \textbf{80}, no.11, 1030 (2020)
[arXiv:2004.03505 [hep-ph]].

  
  \bibitem{ATLAS:2019jst} 
  The ATLAS collaboration [ATLAS Collaboration],
  ATLAS-CONF-2019-029.

\end{thebibliography}
\end{document}